\documentclass[11pt,a4paper]{article}
\usepackage{amsmath}
\usepackage{cite}
\begin{document}
\title{On the relation between Airy integral and Bessel functions revisited}
\author{Mehdi Tabrizi\footnote{tabrizi@bk.ru, tabrizi@razi.ac.ir}\,\,\footnote{http://fm.razi.ac.ir/tabrizi}\,,\,
Ebrahim Maleki Harsini \\ 
\\
\small \em Department of Physics, Faculty of Science, \\
\small \em Razi University of  Kermanshah, 67149-67346, Iran}
\date{}
\maketitle
\begin{abstract}
The Airy integral and Bessel functions are of significant in mathematical description of spectral distribution of different types of radiation produced by relativistic charged particles moving in Synchrotron and in periodical micro- and macro-structures. A simple proof of the relation between Airy integral and modified Bessel function is of most important for radiation physicists, who are not expert in pure mathematics. In this paper we discuss an old proof proposed by Nicholson, and suggest another simple one based on the Bowman transformation of Bessel differential equation with a complex constant.
\end{abstract}
\section{Introduction}
Special functions play a fundamental role in radiation physics. Of special interest are Airy integral and modified Bessel function which appear in spectral distribution of, say, synchrotron radiation\cite{shulga,baier,hofmann}. Many researchers involved with theory of radiation produced by relativistic electrons moving in Synchrotron and insertion devices, such as Undulator and Wiggler\cite{hofmann}, encounter mathematical difficulties  \textit{proving} relations between Airy integral and Bessel functions. Two of such relations 
presented in textbooks\footnote{See for example page 678 in \cite{jackson}.},
which are usually used in calculation of spectral distribution of radiation are
\begin{equation}
\label{A0}
\int_{0}^{\infty}\cos\left[\frac{3}{2}\xi \left(x+\frac{1}{3} x^{3} \right) \right] dx=\frac{1}{\sqrt{3}}K_{1/3}\left(\xi \right)\,\,,
\end{equation}	
and its derivative
\begin{equation}
\label{A01}
\int_{0}^{\infty} x\sin\left[\frac{3}{2}\xi \left(x+\frac{1}{3} x^{3} \right) \right] dx=\frac{1}{\sqrt{3}}K_{2/3}\left(\xi \right)\,.
\end{equation}
By changing the independent variable, left-hand side of relation (\ref{A0}) can be transformed to the Airy integral $\int_{0}^{\infty}\cos(\omega^{3}+\rho \omega)d\omega$, which appeared in another form in researches of Airy\cite{airy} and was tabulated by him. To evaluate the Airy integral, De Morgan obtained a series in ascending powers of $\rho$, which was communicated to Airy \cite{airy1}. His result can be expressed in terms of Bessel functions. Another method to evaluate Airy integral was proposed by Jeffreys\cite{jeff}, who found approximate solutions of certain types of differential equations. 
His solution was related directly to the Airy integral. Of course, by analytic continuation and integration in complex plane applying Jordan's lemma and Cauchy's theorem, one can evaluate the Airy integral expressing result in terms of Bessel function (see for example \cite{watson,olver}) in order to prove relations (\ref{A0}), 
but all aforementioned mathematical processes are excessively laborious (see also \cite{brill}).\\
Unfortunately, physicists have to pass through an abundance of papers and books on special functions to find a simple method for non-experts in pure mathematics to prove such relations.
Authors of this paper could find just one book\cite{olivier}, in which the Nicholson's work\cite{Nich} was cited. It was shown in his paper that Airy integral satisfying  an identical differential equation with the normal form of equation for Bessel function $I$ of order $\frac{1}{3}$ can be expressed in terms of the latter\footnote{There is a misprint of sign in equation (11) of Nicholson's work\cite{Nich}.}. Nicholson's method of evaluation of Airy integral is more convenient for physicists, since the Bessel differential equation is of significant in mathematical description of many physical phenomena and physicists are familiar with it. Although Nicholson found a special case of Bessel differential equation 
in order to express Airy integral in terms of function $I$, it is interesting for physicists to prove relation (\ref{A0}) using another transformation of Bessel differential equation.\\
In this paper we show that there exists a more general form of identical differential equation of the second order with the form of obtained equation by Nicholson. The differential equation can be transformed to the Bessel equation 
with appropriate choosing of numerical value of its constants.
The result is expressed\footnote{This is not done in Nicholson's paper, but is written in textbooks and Handbooks\cite{bateman,abram,grad}.} in terms of modified Bessel function $K$ of order $\frac{1}{3}$.
\\
The paper is organized as follows. In \S\ref{cal} we briefly discuss the Nicholson's method to find a differential equation, which the Airy integral satisfies. On the basis of Bowman representation, we present in \S\ref{bowm} an identical differential equation with the Nicholson's result, then we give a simple proof of relation (\ref{A01}). The paper is concluded with summary in \S\ref{summ}.
\section{Nicholson's method}\label{cal}
Following Nicholson in this section, we discuss in brief his method to find relation between Airy integral and Bessel function $I$ of order $\frac{1}{3}$. 
The left-hand side of relation (\ref{A0}) is a function of $\xi$. 
Changing independent variable $x$ to new $\omega$ as $\frac{1}{2} \xi x^{3} =\omega^{3}$, 
we rewrite 
integral in relation (\ref{A0}) as follows
\begin{equation}
 \int_{0}^{\infty}\cos\left[\frac{3}{2}\xi \left(x+\frac{1}{3} x^{3} \right) \right] dx=\left(\frac{2}{\xi} \right) ^{\frac{1}{3}}\int_{0}^{\infty}\cos\left[ \omega^{3} +\rho\omega\right]  d\omega\,\,,
\label{A11}
\end{equation}
where $\rho\equiv3\left(\frac{\xi}{2} \right) ^{\frac{2}{3}}$ is a new parameter.\\
Integral $A(\rho)=\int_{0}^{\infty}\cos(\omega^{3}+\rho \omega)d\omega$ in the right-hand side of equation (\ref{A11}) is the one, which Nicholson started to work with.
One of the methods to evaluate $A(\rho)$ is to find a differential equation, which $A(\rho)$ satisfies\footnote{In what follows we use accents for differentiation with respect to appropriate variable.}.\\
Following Nicholson,  we consider two functions $f\left(\rho \right)=\int_{0}^{\infty}\cos\left(\omega^{3}+\rho\omega \right) d\omega$ and $ g\left(\rho \right)=\int_{0}^{\infty}\sin\left(\omega^{3}+\rho\omega \right) d\omega$.
Defining 
$u(\rho)\equiv\int_{0}^{\infty} e^{-i\left( \omega^{3}+\rho\omega\right) }d\omega$, 
one can express it as $u(\rho)=f\left(\rho \right) -i g\left(\rho \right)$\footnote{We are interested in this paper in $\Re u(\rho)$.}.
Differentiating two times integral representation of $u(\rho)$ with respect to $\rho$, 
one finds that $u(\rho)$ satisfies equation $u^{''}-\frac{1}{3}\rho u=\frac{1}{3}i$. 
Putting in the latter $ f\left(\rho \right) $ and $ g\left(\rho \right)$ instead of $u(\rho)$, 
one comes to the differential equation in terms of $f(\rho)$ and $g(\rho)$. 
Isolating real and imaginary parts of the latter, 
we have for real part
\begin{equation}
\label{A112}
f^{''}-\frac{1}{3} \rho f=0\,\,,
\end{equation}
which is called the Airy differential equation, and  was considered by Jeffreys\cite{jeff}.\\
Changing $f(\rho)$ as $f(\rho)=\rho^{-1/4}F(\rho)$ and putting it into equation (\ref{A112}), 
one can rewrite the latter in ascending powers of $\rho$
\begin{equation}
\label{A113}
\frac{5}{16}\rho^{-9/4}F - \frac{1}{2}\rho^{-5/4}F^{'} +\rho^{-1/4}F^{''}- \frac{1}{3}\rho^{3/4}F=0\,\,.
\end{equation}
To reduce equation (\ref{A113}) to a familiar one, we 
first multiply it 
by $ \rho^{1/4} $, then we change independent variable $\rho$ to new $y$ as $ \rho^{3}=y^{2} $, and rewrite equation (\ref{A113}) in new variable. In this case we have
\begin{equation}
\label{eq12c}
F^{''} + \left( \frac{5}{36y^{2}}-\frac{4}{27}\right)F=0 \,\,.
\end{equation}
Equation (\ref{eq12c}) is the one, which Nicholson obtained \cite{Nich}, and is a special case of Bessel differential equation $x^{2}y^{''}+xy^{'}+(x^{2}-\nu^{2})y=0$. The latter by substitution $y=ux^{-1/2}$ is reduced to $u^{''}+\left[1+(1-4\nu^{2})/4x^{2}\right]u=0$, which is the same as eqaution (\ref{eq12c}), provided that $\frac{5}{36}=\frac{1}{4}-\nu^{2}$ or $\nu=\pm1/3$. 
Returning back to equation (\ref{A112}), one can express general solution for $f$ (the Airy integral) in terms of $I_{1/3}$ and $I_{-1/3}$.
Putting $\rho=0$ thereafter in obtained result for $f$ 
and using properties of Gamma function, one can find numerical values of coefficient in general solution for $f$. 
The final result is therefore 
\begin{equation}
\label{eq12c1}
 \int_{0}^{\infty}\cos\left(\omega^{3}+\rho\omega \right) d\omega =\overbrace{\sqrt{\frac{\rho}{3}}\,\,\frac{\pi}{3}\left\lbrace  I_{-1/3}\left( \frac{2\rho}{3} \sqrt{\frac{\rho}{3}}\right)-I_{1/3}\left(\frac{2\rho}{3} \sqrt{\frac{\rho}{3}}\right)\right\rbrace}^{G}\,.
\end{equation}
Relation (\ref{eq12c1}) was obtained on the basis of solution of differential equation (\ref{eq12c}), and
is written in References and Handbooks\cite{watson,abram}, but it differs in the signs of $I_{-1/3}$ and $I_{1/3}$ from the result obtained by Nicholson (equation (11) in page 9 of \cite{Nich}). We have designated right-hand side of relation (\ref{eq12c1}) by $G$, which will be discussed in the next section.
\section{Bowman transformation of Bessel equation}\label{bowm}
Nicholson obtained 
the particular case of Bessel differential equation. The question is whether there exists another transformation of Bessel equation with a complex constant, which is identical to the equation (\ref{eq12c}).\\
In 1958, Bowman introduced a general differential equation of second order\cite{bowman}, which can be written in modern notations as follows
\begin{equation}\label{eq13c}
x^{2}y^{''} + \left(1-2c \right) xy^{'} + \left( b^{2}a^{2}x^{2a} + c^{2} - \nu^{2}a^{2}\right)y=0\,,
\end{equation}
where $a$, $b$, $c$, $\nu$ are constants, and $x$ is independent variable.\\
He considered particular cases of equation (\ref{eq13c}) by giving different real, but not complex, values to the constants $a$, $b$, $c$, $\nu$. He wrote that solutions of such equations can at once be written in terms of Bessel functions.\\
In fact, equation (\ref{eq13c}) by changing variables in two subsequent steps can be transformed to Bessel differential equation\cite{bell}. Replacing $y$ as $y=x^{c}u$ and putting it into equation (\ref{eq13c}), one comes to $x^{2}u^{''} + xu^{'} + a^{2}\left(b^{2}x^{2a} - \nu^{2} \right) u=0$, 
and subsequent changing of variable $z=bx^{a}$ gives
$z^{2}u^{''} + zu^{'} + \left(z^{2} - \nu^{2} \right) u=0$. 
The latter 
 is the Bessel differential equation. If $\nu$ is not an integer, then solution can be expressed in terms of $J$ of order $\pm\nu$ and of a real argument\cite{bowman}.\\
Bowman solved equation (\ref{eq13c}) for real constants. To come to equation (\ref{eq12c}), we consider a complex value for  $b$, namely, we give following values for constants in equation (\ref{eq13c}): $a=1$, $b=\frac{2i}{3\sqrt{3}}$, $c=\frac{1}{2}$ and $\nu=\frac{1}{3}$. Equation (\ref{eq13c}) in this case is the same as equation (\ref{eq12c}), but with solution in terms of Bessel function $J$ of order $\frac{1}{3}$ and of an imaginary argument: $J_{\pm\frac{1}{3}\nu}(iu)$, where $u\equiv\left(\frac{2}{3\sqrt{3}}y \right)$. Using relation between Bessel functions $J$ of an imaginary argument and $I$ of a real argument $I_{1/3}(u)=i^{-1/3}J_{1/3}(iu)$, one can express solution of equation (\ref{eq13c}) with a complex constant and non-integer $\nu$ in terms of $I_{1/3}(u)$. Therefore, using chosen values for constants and returning back from Bowman transformation of Bessel equation (\ref{eq13c}) to the Airy differential equation (\ref{A112}), one can evaluate Airy integral and come to the result (\ref{eq12c1}).\\
Function $G$ in the right-hand side of relation (\ref{eq12c1}) is less suitable for practical application in radiation physics. Therefore, one can use relation $K_{\nu}(x)=\frac{\pi}{2\sin\pi\nu}\left[I_{-\nu}(x)-I_{\nu}(x)\right]$ in order to express final result in terms of modified Bessel function. It is found at once that 
\begin{equation}
\label{eq13c2}
 \int_{0}^{\infty}\cos\left(\omega^{3}+\rho\omega \right) d\omega=\frac{\sqrt{\rho}}{3} K_{1/3}\left(\frac{2\rho}{3} \sqrt{\frac{\rho}{3}} \right)\,.
\end{equation}
Returning thereafter back from $\rho$ to $\xi$ using $\rho\equiv3\left(\frac{\xi}{2} \right) ^{\frac{2}{3}}$ and putting right-hand side of relation (\ref{eq13c2}) in the right-hand side integral of equation (\ref{A11}), one comes to relation between Airy integral and modified Bessel function (\ref{A0}).\\
At the end of this section we test by a simple method validity of relation (\ref{A01}), which can be derived from relation (\ref{A0}) by differentiation of the latter and applying properties of modified Bessel function $K$. For simplicity we first differentiate left-hand side of relation (\ref{A0}) using equation (\ref{A11}). In this case we have
\begin{equation}
\label{eq13c3}
\begin{split}
&\frac{d}{d\xi}\left[\int_{0}^{\infty}\cos\left[\frac{3}{2}\xi \left(x+\frac{1}{3} x^{3} \right) \right] dx\right]=\\&
- \frac{1}{3\sqrt{3}\,\xi}K_{1/3}\left(\xi \right) - \int_{0}^{\infty}x\sin\left[\frac{3}{2}\xi \left(x+\frac{1}{3} x^{3} \right) \right] dx\,.
\end{split}
\end{equation}
Differentiating right-hand side of relation (\ref{A0}), and applying recurrence relations $K_{\nu -1}\left(\xi \right) + K_{\nu +1}\left(\xi \right)=-2K_{\nu}^{'}\left( \xi\right)$ and  $K_{\nu -1}\left(\xi \right) - K_{\nu +1}\left(\xi \right)=-\frac{2\nu}{\xi}K_{\nu}\left( \xi\right)$, one comes to the next result
\begin{equation}
\label{eq13c4}
\frac{1}{\sqrt{3}}K^{'}_{1/3}\left(\xi \right)=\frac{1}{\sqrt{3}}\left[ - K_{-2/3}\left(\xi \right)-\frac{1}{3\xi}K_{1/3}\left( \xi\right)\right].
\end{equation}
Setting relation (\ref{eq13c4}) equal to relation (\ref{eq13c3}) and taking into account that $K_{\nu}(\xi)=K_{-\nu}(\xi)$, one comes to the final result 
\begin{equation}\label{eq13c5}
\int_{0}^{\infty}x\sin\left[\frac{3}{2}\xi \left(x+\frac{1}{3} x^{3} \right) \right] dx = \frac{1}{\sqrt{3}} K_{2/3}\left(\xi \right)\,, 
\end{equation} 
which together with relation (\ref{A0}) are used nowadays in radiation physics\cite{hofmann,duke}.
\section{Summary}\label{summ}
The Nicholson's method to relate Airy integral to Bessel modified function was revisited. We shown that equation obtained by Nicholson is a particular case of a more general transformation of Bessel equation proposed by Bowman. We shown that in addition to real values of constants in Bowman transformation a complex value for one of the constants can be chosen in order to obtain an equation identical with the form obtained by Nicholson. In the latter case Airy integral can be express in terms of original Bessel $J$ function of non-integer order and of an imaginary argument, which in turn can be expressed in terms of modified Bessel function $K$.\\
A simple method for non-experts in pure mathematics was applied to prove relation (\ref{A01}).
\bibliography{mybibfile}

\end{document}